\documentclass[conference]{IEEEtran}
\IEEEoverridecommandlockouts
\usepackage{bm}
\usepackage{cite}
\usepackage{amsmath,amssymb,amsfonts}
\usepackage{algorithmic}
\usepackage{graphicx}
\usepackage{textcomp}
\usepackage{epsfig}
\usepackage{times}
\usepackage{longtable}
\usepackage{supertabular,booktabs}
\usepackage{ltablex}
\usepackage{array}
\usepackage{setspace}
\usepackage{xcolor,soul}
\usepackage{balance}
\usepackage{etoolbox}
\usepackage[labelsep=period,font={small}]{caption}

\newcolumntype{C}[1]{>{\centering\arraybackslash} m{#1}}

\def\BibTeX{{\rm B\kern-.05em{\sc i\kern-.025em b}\kern-.08em
    T\kern-.1667em\lower.7ex\hbox{E}\kern-.125emX}}

\newcommand{\qed}{\hfill $\square$}

\renewcommand{\thetable}{\arabic{table}}

\begin{document}

\title{\LARGE \bf PCT-Based Trajectory Tracking for Underactuated Marine Vessels}

\author{Ji-Hong Li
\thanks{*This work was partially supported by the Korea Institute of Marine Science \& Technology Promotion (KIMST) funded by the Ministry of Ocean and Fisheries (RS-2024-00432366), also in part by the same organization with the grant No. RS-2023-00256122, all in the Republic of Korea.}
\thanks{J. H. Li is with the Autonomous Systems R\&D Division, Korea Institute of Robotics and Technology Convergence,
        Jigok-Ro 39, Nam-Gu, Pohang 37666, Republic of Korea. {\tt\small jhli5@kiro.re.kr}}%
}

\maketitle

\begin{abstract}
This paper investigates the trajectory tracking problem of underactuated marine vessels within a polar coordinate framework. By introducing two polar coordinate transformations (PCTs), the original two-input-three-output second-order tracking model expressed in the Cartesian frame is reduced to a two-input-two-output feedback system. However, the resulting model does not necessarily satisfy the strict-feedback condition required by conventional backstepping approaches. To circumvent potential singularities arising in the controller design, a novel concept termed exponential modification of orientation (EMO) is proposed. While the PCTs yield substantial structural simplification, they also introduce inherent limitations, most notably singularities associated with angular coordinates. Addressing these singularities constitutes another key focus of this paper. Numerical simulation results are presented to demonstrate the effectiveness of the proposed control strategy.
\end{abstract}

\section{Introduction}
Marine vessels are inherently underactuated systems: only two control inputs are available to regulate motion in three degrees of freedom (3-DOF), which poses substantial challenges in controller design. This limitation has motivated extensive research over several decades. Existing studies--primarily distinguished by how reference trajectories are formulated--can be broadly classified into trajectory tracking \cite{b1}--\hspace{1sp}\cite{b13} and path following \cite{b14}--\hspace{1sp}\cite{b16}. Unlike trajectory tracking, which relies on time-parameterized references, path following replaces time with a path parameter that is often treated as an auxiliary control input. Proper design of this variable enables Lyapunov-based stabilization and affords additional flexibility in controller design \cite{b14}--\hspace{1sp}\cite{b17}.

The trajectory tracking literature \cite{b1}--\hspace{1sp}\cite{b13} can be further divided into two categories. The first aims to track all states of a predefined 3-DOF reference trajectory \cite{b1}--\hspace{1sp}\cite{b7}. The second considers reduced tracking objectives motivated by the limitation of having only two control inputs. For example, \cite{b8,b9,b12} guide the vessel to follow a moving point along the reference trajectory while ignoring the reference yaw angle. Related formulations using suitable variable transformations appear in \cite{b11,b13}. In \cite{b10}, the reference yaw angle is redefined as the azimuth angle from the vessel to the moving target point, making it state-dependent rather than predefined. More recent studies \cite{b18,b19} continue to develop methods in this direction. For clarity, the first category is referred to in this paper as traditional trajectory tracking.

This paper focuses on traditional trajectory tracking. When the vessel dynamics are minimum phase \cite{b8}, certain position-coordinate transformations \cite{b1}--\hspace{1sp}\cite{b3} or yaw-angle error transformations \cite{b4} reduce the system to a triangular-like structure, allowing the use of integrator backstepping \cite{b1}. In the non-minimum phase case, the coordinate shift proposed in \cite{b9} selects a point in the body-fixed frame such that the rudder action induces pure rotation without sway, thereby restoring the minimum-phase property. These approaches \cite{b1}--\hspace{1sp}\cite{b4} rely on simplified vessel dynamics and Lyapunov direct method design. To guarantee stability, several restrictions on the reference trajectory were introduced, including persistency-of-excitation conditions on the yaw rate \cite{b1}--\hspace{1sp}\cite{b3}, which were later relaxed in \cite{b4}. However, none of these works provides a unified controller applicable to both zero and nonzero yaw-rate cases. In addition, some gain-selection conditions remain difficult to satisfy in practice \cite{b1,b2}. Although \cite{b7} considers a more general vessel model with a nonlinear sliding-mode approach, it also does not unify these cases.

To address these limitations, \cite{b6} proposed a new control framework for underactuated marine vessels with general dynamics and bounded uncertainties. Unlike earlier methods, it imposed no restrictions on the reference trajectory. By applying two polar coordinate transformations (PCTs), the tracking model is reduced to a second-order two-input-two-output system, enabling the use of a general backstepping method \cite{b20}. However, the use of polar coordinates introduces additional challenges. First, the reduced system does not necessarily satisfy the strict-feedback condition, which may lead to singularities in recursive controller design. To address this issue, \cite{b6} introduced the asymptotic modification of orientation (AMO) technique, later adopted in subsequent studies for both 2D \cite{b21,b22} and 3D extensions \cite{b23,b24}. Second, polar coordinates inherently introduce a singularity due to the undefined polar angle at the origin. To avoid this difficulty, all related works \cite{b6},\cite{b21}--\hspace{1sp}\cite{b24} assume that the vessel's surge speed remains strictly positive along the entire trajectory.

Meanwhile, control barrier functions (CBFs) \cite{b25}--\cite{b27} have recently been widely applied not only for system safety enforcement but also for singularity avoidance in robotics and control. With appropriately designed CBFs, various practical singularities can be effectively prevented.

Motivated by these observations, this paper proposes a new trajectory tracking scheme for underactuated marine vessels with the following main contributions:
\begin{itemize}
  \item The AMO framework developed in \cite{b6,b21,b22} is extended to an exponential modification of orientation (EMO) formulation, which guarantees exponential--rather than merely asymptotic--stability of the closed-loop tracking system in the absence of uncertainties.
  \item The restrictive assumption of strictly positive surge speed imposed in \cite{b6},\cite{b21}--\hspace{1sp}\cite{b24} is revisited from two perspectives. First, it is relaxed by requiring instead that the vessel's sway speed be bounded by a known constant, a condition typically satisfied in practical marine vessels. Second, the assumption is completely removed by incorporating a CBF-based constraint. While both approaches significantly relax the original condition, the latter eliminates it entirely at the expense of introducing additional constraints to accommodate uncertainties in the vessel dynamics.
\end{itemize}

The remainder of the paper is organized as follows. Section II presents the vessel kinematic and dynamic models, reformulated into a two-input-two-output second-order feedback form via two PCTs, and discusses the associated singularity issues. Section III develops the EMO-based trajectory tracking scheme under a strictly positive surge-speed assumption, which is relaxed in Section IV. Simulation results are presented in Section V, followed by concluding remarks in Section VI.

\section{Problem Statement}
\subsection{Vessel Model}
This paper considers the vessel's kinematics and dynamics as following \cite{b6,b28}
\begin{align}
\dot{\bm{\eta}}&=C_b^n\bm{\nu}, \label{eq1} \\
\dot{\bm{\nu}}&=\bm{f}+\bm{B}\bm{\tau}+\bm{d}, \label{eq2}
\end{align}
where $\bm{\eta}=[x,y,\psi]^T$ with $(x,y)$ denoting the horizontal position and $\psi$ the yaw angle in the navigation frame; $\bm{\nu}=[u,v,r]^T$, where $u$ and $v$ represent the surge and sway velocities, and $r$ is the yaw rate in the body-fixed frame; $\bm{f}=[f_u,f_v,f_r]^T$, where $f_u$, $f_v$, and $f_r$ correspond to the modeled nonlinear dynamics of the vessel, incorporating hydrodynamic damping, inertia (including added mass), Coriolis and centripetal, and gravitational effects in the surge, sway, and yaw directions, each assumed to be of class $C^2$; $\bm{\tau}=[\tau_u,\tau_r]^T$, where the surge force $\tau_u$ and the yaw moment $\tau_r$ are the only available control inputs; $\bm{d}=[d_u,d_v,d_r]^T$ denotes the uncertainty terms, which include model errors, measurement noises, exogenous disturbances, all of which are unmatched \cite{b29}; and the coordinate transformation matrix $\bm{C}_b^n$ from the body-fixed frame to the navigation frame and the control gain matrix $\bm{B}$ are defined as follows:
\begin{equation*}
\bm{C}_b^n=\begin{bmatrix}\cos\psi&-\sin\psi&0\\ \sin\psi&\cos\psi&0\\0&0&1\end{bmatrix},~~~\bm{B}=\begin{bmatrix}b_u&0\\0&\epsilon_r\\0&b_r\end{bmatrix},
\end{equation*}
where $b_u,b_r>0$ are constant control gains, and $\epsilon_r$ represents the lift effect induced by the yaw moment \cite{b8}.

\emph{Remark 1}. In (\ref{eq2}), the system is minimum phase when $\epsilon_r=0$, and non-minimum phase otherwise\cite{b8}. This paper considers both cases. Notably, the computability of the proposed tracking method is unaffected by whether the system is minimum phase or not.

Regarding the uncertainty term $\bm{d}=[d_u,d_v,d_r]^T$, this paper considers only the unmatched components, which are assumed to satisfy the following conditions.

\emph{Assumption 1}. The uncertainties are bounded in magnitude by $|d_u|\leq d_{uM}$, $|d_v|\leq d_{vM}$, and $|d_r|\leq d_{rM}$ with $d_{uM}$, $d_{vM}$, $d_{rM}>0$ known positive constants.

\subsection{Two PCTs}
\begin{figure}[!t]
\centerline{\includegraphics[width=5.0cm]{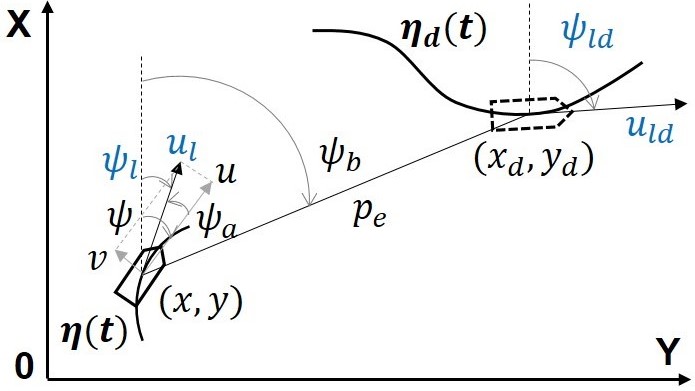}}
\caption{Illustration of two polar coordinate systems and related variables.}
\label{fig1}
\end{figure}

A closer look at the vessel's dynamics in (\ref{eq2}) reveals that the primary control challenge lies in the inability to effectively handle the sway dynamics. With this in mind, this paper first introduces the following PCT in the vessel's body-fixed frame (see Fig. \ref{fig1}):
\begin{equation}
\begin{bmatrix}u_l\\ \psi_a\end{bmatrix}=\mathcal{F}_a(u,v):=\begin{bmatrix}\sqrt{u^2+v^2}\\ \arctan(v/u)\end{bmatrix}, \label{eq3}
\end{equation}
where $\psi_a$ is commonly known as the sideslip angle \cite{b28}.

Using (\ref{eq3}), the vessel's two-input-three-output dynamics (\ref{eq2}) can be reformulated as following
\begin{equation}
\begin{bmatrix}{\dot{u}}_l \\ \dot{r}_l\end{bmatrix}=\begin{bmatrix}f_{u_l} \\ f_{r_l}\end{bmatrix}+\begin{bmatrix}b_{u_l} &\epsilon_{ra} \\0 &b_r\end{bmatrix}\begin{bmatrix}\tau_u\\ \tau_r\end{bmatrix} +\begin{bmatrix}d_{u_l} \\d_r\end{bmatrix}, \label{eq4}
\end{equation}
where $r_l=r+\dot{\psi}_a$, $f_{u_l}=\cos\psi_a f_u+\sin\psi_a f_v$, $f_{r_l}=f_r+\ddot{\psi}_a$, $b_{u_l}=\cos\psi_a b_u$, $\epsilon_{ra}=\sin\psi_a\epsilon_r$, and $d_{u_l}=\cos\psi_a d_u+\sin\psi_a d_v$. According to Assumption 1, it is straightforward to derive that $|d_{u_l}|\leq d_{u_lM}=\sqrt{d_{uM}^2+d_{vM}^2}$.

\emph{Remark 2}. As defined in (\ref{eq3}), the sideslip angle $\psi_a$ becomes either undefined or equal to $\pm\pi/2$ when $u=0$. Consequently, under this condition, the control gain $b_{u_l}$ in (\ref{eq4}) also becomes either undefined or zero. To avoid such singularities, it is often necessary to ensure that $u(t)>0$ holds for all $t\geq0$. From a practical standpoint, most marine vessels control yaw motion via a rudder, with the corresponding control input $\tau_r$ typically being proportional to the square of the vessel's surge speed \cite{b30}. As a result, treating $\tau_u$ and $\tau_r$ as approximately independent control inputs requires the vessel to maintain a sufficiently high forward velocity. Based on these considerations, the condition $u(t)>0$, for all $t\geq0$ was imposed as an assumption in all of the related previous works \cite{b6,b10,b21}--\hspace{1sp}\cite{b24}. In this paper, however, we aim to eliminate this assumption by rigorously analyzing the conditions under which $u(t)>0$ can be guaranteed.

In the context of implementing the vessel's trajectory $\bm{\eta}(t)$, one straightforward approach is to realize it through the vessel's kinematics given in (\ref{eq1}) by specifying $u$, $v$, and $r$. However, in this case, $u$, $v$, and $r$ cannot be arbitrarily assigned, as they must satisfy a nonlinear constraint stemming from the vessel's sway dynamics. An alternative approach is to generates $\bm{\eta}(t)$ using $u_l$ and $\psi_l$, where $u_l$ is defined in (\ref{eq3}), and $\psi_l=\psi+\psi_a$, where $\dot{\psi}_l=r_l$ with $r_l$ defined in (\ref{eq4}), represents the yaw angle of $u_l$ in the navigation frame. The corresponding kinematics is given by
\begin{equation}
\dot{x}=u_l\cos\psi_l, ~~\dot{y}=u_l\sin\psi_l. \label{eq5}
\end{equation}
Under this formulation, $u_l$ and $\psi_l$ can be freely chosen, making trajectory generation significantly more flexible and convenient. In this study, we adopt this latter approach to represent the vessel's trajectory.

Consider a reference trajectory $\bm{\eta}_{ld}(t)=[x_d(t),y_d(t)$, $\psi_{ld}(t)]^T$, where $\dot{x}_d=u_{ld}\cos\psi_{ld}$ and $\dot{y}_d=u_{ld}\sin\psi_{ld}$. The position error is defined as $p_e(t)=||\bm{p}_d(t)-\bm{p}(t)||_2$ with $\bm{p}(t)=[x(t),y(t)]^T$. Now we introduce the following second polar coordinate $(p_e,\psi_b)$ in the navigation frame (see Fig. \ref{fig1}),
\begin{equation}
\begin{bmatrix}p_e\\ \psi_b\end{bmatrix}=\mathcal{F}_b(x_e,y_e):=\begin{bmatrix}\sqrt{x_e^2+y_e^2}\\ \text{atan2}(y_e,x_e)\end{bmatrix}, \label{eq6}
\end{equation}
where $x_e=x_d-x$, $y_e=y_d-y$, and $\psi_b$ is defined as the azimuth angle from the vehicle to the target point moving on the reference trajectory $\bm{\eta}_{ld}(t)$.

According to (\ref{eq6}), the time derivative of $p_e(t)$ becomes
\begin{equation}
\dot{p}_e=u_{ld}cos(\psi_{ld}-\psi_b)-u_l cos(\psi_l -\psi_b). \label{eq7}
\end{equation}

If we define $\psi_{le}=\psi_{ld}-\psi_l$, then, in conjunction with (\ref{eq4}), the trajectory tracking error model can be expressed in the following two-input-two-output second-order feedback form:
\begin{align}
\begin{bmatrix}\dot{p}_e\\ \dot{\psi}_{le}\end{bmatrix}\!&=\!\begin{bmatrix}u_{ld}\cos(\psi_{ld}\!-\!\psi_b)\\ \dot{\psi}_{ld} \end{bmatrix}\!-\!\begin{bmatrix}\cos(\psi_l\!-\!\psi_b) &0\\0 &1\end{bmatrix}\!\!\begin{bmatrix}u_l \\r_l\end{bmatrix}, \label{eq8} \\
\begin{bmatrix}{\dot{u}}_l \\ \dot{r}_l\end{bmatrix}\!&=\!\begin{bmatrix}f_{u_l} \\ f_{r_l}\end{bmatrix}+\begin{bmatrix}b_{u_l} &\epsilon_{ra} \\0 &b_r\end{bmatrix}\begin{bmatrix}\tau_u\\ \tau_r\end{bmatrix} +\begin{bmatrix}d_{u_l} \\d_r\end{bmatrix}. \label{eq9}
\end{align}

In line with the considerations in Remark 2, we impose the following restriction on the reference trajectory.

\emph{Assumption 2}. The reference trajectory $\bm{\eta}_{ld}(t)$ satisfies $u_{ld}(t)\geq u_m>0$, for all $t\geq 0$ with $u_m$ a design parameter.

\subsection{System Singularities}
\subsubsection{$\cos(\psi_l-\psi_b)=0$ and $b_{u_l}=0$}
Recalling the reduced tracking model in (\ref{eq8}) and (\ref{eq9}), singularities may arise in the recursive controller design when $\cos(\psi_l-\psi_b)=0$ or $b_{u_l}=0$, rendering the virtual input $u_l$ and the actual input $\tau_u$ ineffective. The singularity associated with $\cos(\psi_l-\psi_b)=0$ has been addressed via the AMO concept \cite{b6}, which has been adopted in subsequent works \cite{b21,b22}. Moreover, sin $b_{u_l}=b_u\cos\psi_a$ with $b_u\neq 0$, the condition $b_{u_l}=0$ corresponds to $u=0$ with $v\neq 0$, which will be discussed later.

\subsubsection{Polar Angle Singularity}
Although polar coordinates provide structural simplification, they inherently introduce singularities, as the polar angle is undefined at the origin. In (\ref{eq8}) and (\ref{eq9}), both $\psi_a$ and $\psi_b$ are undefined at $u_l=0$ and $p_e=0$. Previous studies \cite{b6,b21,b22} showed that, when $p_e=0$, all terms involving $\psi_b$ vanish, allowing its singularity to be safely disregarded. In contrast, the singularity of $\psi_a$ is critical to the control methods proposed in \cite{b6,b21,b22}. To avoid this issue, as well as the singularity at $\psi_a=\pm\pi/2$, the condition $u(t)>0$\footnote{The case $u<0$ is neglected, as it is incompatible with most practical applications.} for all $t\geq 0$ is imposed. However, this assumption inevitably compromises the theoretical rigor of the control design.

\subsection{Problem Formulation}
In this paper, the tracking problem for (\ref{eq8}) and (\ref{eq9}) is addressed using a general backstepping framework \cite{b20}. The potential singularity arising from $\cos(\psi_l-\psi_b)=0$ is avoided through the introduction of the EMO concept. With respect to the singularities arising from $b_{u_l}=0$ and from the fact that $\psi_a$ is undefined at $u_l=0$, this paper revisits the assumption adopted in \cite{b6},\cite{b21}--\hspace{1sp}\cite{b24} and relaxes it in two different ways. One approach replaces the original assumption with a less restrictive condition, while the other eliminates it altogether by incorporating a CBF-based technique. The respective advantages and limitations of these two approaches are also systematically compared and analyzed.

\section{EMO-Based Tracking Controller}
This section presents the proposed tracking controller under the assumption that $u(t)>0$, for all $t\geq0$; the enforcement of this condition is discussed in detail in the next section.

\subsection{Exponential Modification of Orientation (EMO)}
As mentioned before, to avoid the potential singularity arising from $\cos(\psi_l-\psi_b)=0$, this paper introduces the following concept.

\emph{Definition 1}. Consider the position error kinematics in (\ref{eq7}). Let the reference trajectory be given by $\bm{\eta}_{ld}=[x_d(t),y_d(t)$, $\psi_{ld}(t)]^T\!\!\in\!\!{\bm\Re}^3$ with $\dot{x}_d\!\!=\!\!u_{ld}\cos\psi_{ld}$, $\dot{y}_d\!\!=\!\!u_{ld}\sin\psi_{ld}$. If there exists a modified orientation $(u_{ld}^m(t)$, $\psi_{ld}^m(t))$ such that applying $(u_l,\psi_l)=(u_{ld}^m,\psi_{ld}^m)$ ensures exponential convergence $p_e(t)\rightarrow 0$, and this further guarantees $(u_{ld}^m(t), \psi_{ld}^m(t))\rightarrow (u_{ld}(t), \psi_{ld}(t))$, then $(u_{ld}^m(t), \psi_{ld}^m(t))$ is called an exponential modification of orientation (EMO) of $(u_{ld}(t),\psi_{ld}(t))$.

The following Lemma shows an example of this EMO.

\emph{Lemma 1}. Given the position error kinematics in (\ref{eq7}), suppose the modified orientation $(u_{ld}^m, \psi_{ld}^m)$ is chosen as
\begin{align}
u_{ld}^m&=u_{ld}+c_u p_e\cos\left[(\psi_{ld}-\psi_b)e^{-c_{\psi}p_e}\right], \label{eq10} \\
\psi_{ld}^m&=\psi_b+(\psi_{ld}-\psi_b)e^{-c_{\psi}p_e}, \label{eq11}
\end{align}
where $c_u,c_{\psi}>0$ are design parameters selected such that $c_{\psi}u_m-2c_u\kappa>0$, where the constant $\kappa$ will be defined later. Then, the pair $(u_{ld}^m, \psi_{ld}^m)$ constitutes an EMO of $(u_{ld},\psi_{ld})$.

\emph{Proof}.
Substituting (\ref{eq10}) and (\ref{eq11}) with $(u_l,\psi_l)=(u_{ld}^m,\psi_{ld}^m)$ into (\ref{eq7}) yields
\begin{align}
\dot{p}_e=&u_{ld}\cos(\psi_{ld}\!-\!\psi_b)\!-\!\left\{u_{ld}\!+\!c_up_e\cos\left[(\psi_{ld}\!-\!\psi_b)e^{-c_{\psi}p_e}\right]\right\} \nonumber \\
&\cos\left[(\psi_{ld}-\psi_b)e^{-c_{\psi}p_e}\right]. \label{eq12}
\end{align}

With (\ref{eq12}), further we can get the following expansion,
\begin{align}
\dfrac{\partial \dot{p}_e}{\partial p_e}&=-c_u\cos^2\varphi-c_{\psi}u_{ld}\varphi\sin\varphi-2c_uc_{\psi}p_e \varphi\sin\varphi \cos\varphi \nonumber \\
&= -c_u\cos^2\varphi-c_{\psi}\varphi\sin\varphi\left(u_{ld}+2c_up_e\cos\varphi\right) \nonumber \\
&\leq -c_u\cos^2\varphi-c_{\psi}\varphi\sin\varphi\left(u_{ld}-2c_u\dfrac{\kappa}{c_{\psi}}\right) \nonumber \\
&\leq -c_u\cos^2\varphi-\left(c_{\psi}u_m-2c_u\kappa\right)\varphi \sin\varphi \nonumber \\
&\leq -c_u\cos^2\varphi-\left(c_{\psi}u_m-2c_u\kappa\right)\sin^2\varphi \nonumber \\
&= -c_u+\left[c_u-\left(c_{\psi}u_m-2c_u\kappa\right)\right] \sin^2\varphi \nonumber \\
&\leq -c , \label{eq13}
\end{align}
where $\varphi=(\psi_{ld}-\psi_b)e^{-c_{\psi}p_e}\in[-\pi,\pi]$, and $c=\text{min}\{c_u$, $c_{\psi}u_m-2c_u\kappa\}$.

From (\ref{eq13}) with $\dot{p}_e=0$ at $p_e=0$, it is obvious that $\dot{p}_e\leq -cp_e$. Now consider the following Lyapunov function candidate
\begin{equation}
V=0.5p_e^2. \label{eq14}
\end{equation}

Differentiating (\ref{eq14}) and using $\dot{p}_e\leq -cp_e$ yields
\begin{equation}
\dot{V}\leq -cp_e^2=-2cV, \label{eq15}
\end{equation}
and this guarantees the exponential convergence $p_e\rightarrow 0$, which, together with (\ref{eq10}) and (\ref{eq11}), ensures $(u_{ld}^m,\psi_{ld}^m)\rightarrow (u_{ld},\psi_{ld})$. \qed

In the Lemma 1, the constant $\kappa$ is defined in the following proposition.

\emph{Proposition 1}. For the function $f(\zeta,\phi)=\zeta\cos\left(\phi e^{-c_{\psi}\zeta}\right)$, defined over $\zeta\geq 0$ and $\phi\in[-\pi,\pi]$, it follows that
\begin{equation}
\min_{\zeta\geq0}f(\zeta,\phi)=-\dfrac{\kappa}{c_{\psi}}\approx -\dfrac{0.2099}{c_{\psi}}. \label{eq16}
\end{equation}

\emph{Proof}. First, for $\phi\in[-\pi,\pi]$, it is easy to verify that $f(\zeta,\phi)\geq f(\zeta,\phi)|_{\phi=\pm\pi}=\zeta \cos\left(\pi e^{-c_{\psi}\zeta}\right)\equiv f(\zeta)$. Moreover, since $f(0)=f(\ln2/c_{\psi})=0$, $f(\zeta)$ attains a minimum (or minima) within the interval $\zeta\in(0,\ln2/c_{\psi})$.

Let $z=\pi e^{-c_{\psi}\zeta}$ so that $z\in(0,\pi]$. Then, $f(z)=-[\ln(z/\pi)/c_{\psi}]\cos z$. Differentiating with respect to $z$, the condition $f'(z)=0$ leads to the transcendental equation $\cos z=z \ln(z/\pi)\sin z$, which has a unique solution $z^*\in(\pi/2,\pi)$; numerically $z^*\approx 2.2253$. Consequently, $\zeta^*=-\ln(z^*/\pi)/c_{\psi}\approx 0.3448/c_{\psi}$ and
\begin{equation}
f(\zeta)_{min}=f(\zeta^*)=\zeta^*\cos\left(\pi e^{-c_{\psi}\zeta^*}\right)\approx -\dfrac{0.2099}{c_{\psi}}. \label{eq17}
\end{equation}
This completes the proof. \qed

\emph{Remark 3}. For a given reference $(u_{ld},\psi_{ld})$, by introducing its EMO $(u_{ld}^m,\psi_{ld}^m)$ and enforcing vessel tracking of this EMO, the potential singularity caused by $\cos(\psi_{ld}-\psi_b)=0$ can be avoided. In this context, $u_{ld}^m$ serves as the stabilizing function for the virtual input $u_l$.

\subsection{Controller Design}
\subsubsection{Kinematic Tracking}
As noted earlier, given the reference trajectory $\bm{\eta}_{ld}$ with $(u_{ld},\psi_{ld})$, the strategy adopted in this paper to avoid potential singularity is to steer the vessel to track the EMO $(u_{ld}^m,\psi_{ld}^m)$ rather than the original $(u_{ld},\psi_{ld})$. Accordingly, the following Lyapunov function candidate is introduced at this step:
\begin{equation}
V_1=0.5\left(p_e^2+\gamma_{\psi}\psi_{le}^2\right), \label{eq18}
\end{equation}
where $\psi_{le}=\psi_{ld}^m-\psi_l$ and $\gamma_{\psi}>0$ is a weighting factor.

Suppose $u_{ld}^m$ and $\psi_{ld}^m$ are chosen as (\ref{eq10}) and (\ref{eq11}). Differentiating (\ref{eq18}) and substituting (\ref{eq8}) yields
\begin{align}
\dot{V}_1=&~p_e\left[u_{ld}\cos(\psi_{ld}-\psi_b )-u_{ld}^m\cos(\psi_{ld}^m-\psi_b)\right] \nonumber \\
&+p_e\left[u_{ld}^m \cos(\psi_{ld}^m-\psi_b)-u_l \cos(\psi_l-\psi_b)\right] \nonumber \\
&+\gamma_{\psi}\psi_{le}\left(\dot{\psi}_{ld}^m-r_l\right) \nonumber \\
\leq&-cp_e^2+p_e\left[u_{le}\cos(\psi_{ld}^m-\psi_b)-2u_l \sin A_{\psi} \sin\frac{\psi_{le}}{2}\right] \nonumber \\
&+\gamma_{\psi}\psi_{le}\left(\dot{\psi}_{ld}^m-\alpha_{r_l}+e_{r_l} \right), \label{eq19}
\end{align}
where $c>0$ is defined in (\ref{eq13}), $u_{le}=u_{ld}^m-u_l$, $A_{\psi}=(\psi_{ld}^m+\psi_l)/2-\psi_b$, and $\alpha_{r_l}$ is a stabilizing function for virtual input $r_l$ and $e_{r_l}=\alpha_{r_l}-r_l$.

In addition to that the stabilizing function for virtual input $u_l$ is taken as $u_{ld}^m$, according to (\ref{eq19}), the remained control law for $\alpha_r$ is chosen as
\begin{equation}
\alpha_{r_l}=\dot{\psi}_{ld}^m+\gamma_{\psi}^{-1}\left[k_{\psi}\psi_{le}-p_eu_l \sin A_{\psi} \frac{\sin\left(\psi_{le}/2\right)}{\psi_{le}/2}\right], \label{eq20}\\
\end{equation}
where $k_{\psi}>0$ is a design parameter.

By substituting (\ref{eq20}) into (\ref{eq19}), we obtain
\begin{equation}
\dot{V}_1\leq -cp_e^2\!-\!k_{\psi}\psi_{le}^2+p_eu_{le}\cos(\psi_{ld}^m-\psi_b)+\gamma_{\psi}\psi_{le}e_{r_l}.  \label{eq21}
\end{equation}

\subsubsection{Dynamic Tracking}
The tracking dynamics (\ref{eq9}) can be rewritten in the following error form:
\begin{equation}
\begin{bmatrix}{\dot{u}}_{le} \\ \dot{e}_{r_l}\end{bmatrix}\!=\!\begin{bmatrix}\dot{u}_{ld}^m-f_{u_l} \\ \dot{\alpha}_{r_l}-f_{r_l}\end{bmatrix}-\overbrace{\begin{bmatrix}b_{u_l} &\epsilon_{ra} \\0 &b_r\end{bmatrix}}^{\textstyle \bm{B}_l}\begin{bmatrix}\tau_u\\ \tau_r\end{bmatrix} -\begin{bmatrix}d_{u_l} \\d_r\end{bmatrix}. \label{eq22}
\end{equation}

Consider the Lyapunov function candidate as follows:
\begin{equation}
V_2=V_1+\dfrac{1}{2}\bm{e}^T\bm{G}\bm{e}, \label{eq23}
\end{equation}
where $\bm{e}=[u_{le},e_{r_l}]^T$, and $\bm{G}=diag(\gamma_u,\gamma_r)$ with $\gamma_u,\gamma_r>0$ weighting factors.

Differentiating (\ref{eq23}) and substituting (\ref{eq22}) and (\ref{eq23}) into the result, we obtain the following expansion
\begin{align}
\dot{V}_2\leq & -c p_e^2-k_{\psi}\psi_{le}^2+p_eu_{le}cos(\psi_{ld}^m-\psi_b)+\gamma_{\psi}\psi_{le}e_{r_l} \nonumber \\
&+\bm{e}^T \bm{G}\left\{\begin{bmatrix}\dot{u}_{ld}^m-f_{u_l}\\ \dot{\alpha}_{r_l}-f_{r_l}\end{bmatrix}-\bm{B}_l\begin{bmatrix}\tau_u\\ \tau_r\end{bmatrix}-\begin{bmatrix}d_{u_l}\\d_r\end{bmatrix} \right\}. \label{eq24}
\end{align}

According to (\ref{eq24}), the final control law is chosen as
\begin{equation*}
\begin{bmatrix}\tau_u\\ \tau_r\end{bmatrix}\!\!=\!\!\bm{B}_l^{-1}\!\!\left\{\!\begin{bmatrix}\dot{u}_{ld}^m\!-\!f_{u_l}\\ \dot{\alpha}_{r_l}\!-\!f_{r_l}\end{bmatrix}\!\!+\!\bm{G}^{-1}\!\left(\bm{K}\bm{e}\!+\!\begin{bmatrix}p_e\cos(\psi_{ld}^m\!-\!\psi_b)\\ \gamma_{\psi}\psi_{le}\end{bmatrix}\right)\right.
\end{equation*}
\begin{equation}
+\left.\!\!\begin{bmatrix}d_{u_lM}\varphi\!(\gamma_u u_{le}d_{u_lM}\!)\\ d_{rM}\varphi\!(\gamma_r e_{r_l}d_{rM}\! )\!\!\end{bmatrix}\!\! \right\}\!\!,~~~~~~~~~~~~~~~~~~~~~~~~ \label{eq25}
\end{equation}
where $\bm{K}=diag(k_u,k_r)$ with $k_u,k_r>0$ design parameters. And $\varphi(\cdot)$ satisfies the following Lemma.

\emph{Lemma 2} (Lemma 1 in \cite{b10}). For any $\epsilon >0$, there exist a smooth function $\varphi(\cdot)$ such that $\varphi(0)=0$, and the following inequality holds:
\begin{equation}
|\zeta|\leq \zeta \varphi(\zeta)+\epsilon,~~\forall \zeta \in \Re. \label{eq26}
\end{equation}

\emph{Remark 4}. Simple examples of functions that satisfy \emph{Lemma 2} include $\varphi(\zeta)=[1/(4\epsilon)]\zeta$, as suggested in \cite{b3}, and $\varphi(\zeta)=tanh(\sigma\zeta/\epsilon)$, where $\sigma=e^{-(\sigma+1)}$, as in \cite{b31}. On the other hand, choosing $\varphi(\zeta)=sgn(\zeta)$ also satisfies inequality (\ref{eq32}) even for $\epsilon=0$. However, due to the discontinuity of the sign function, this choice may lead to chattering issue in practical applications.

Substituting (\ref{eq25}) into (\ref{eq24}) yields
\begin{align}
\dot{V}_2\leq & -cp_e^2-k_{\psi}\psi_{le}^2-k_uu_{le}^2-k_re_{r_l}^2+[|u_{le}|,|e_{r_l}|] \nonumber \\
&~\bm{G}\begin{bmatrix}d_{u_lM}\\ d_{rM}\end{bmatrix}-\bm{e}^T\bm{G}\begin{bmatrix}d_{u_lM}\varphi(\gamma_u u_{le}d_{u_lM}) \\ d_{rM}\varphi(\gamma_r e_{r_l}d_{rM})\end{bmatrix} \nonumber \\
\leq & -cp_e^2-k_{\psi}\psi_{le}^2-k_uu_{le}^2-k_re_{r_l}^2+\epsilon_{u_l}+\epsilon_{r_l} \nonumber \\
\leq &-\lambda V_2+\varepsilon, \label{eq27}
\end{align}
where $\lambda:=\text{min}\{2c,2k_{\psi}\gamma_{\psi}^{-1},2k_u\gamma_u^{-1},2k_r\gamma_r^{-1}\}$, and $\varepsilon=\epsilon_{u_l}+\epsilon_{r_l}$ with $\epsilon_{u_l},\epsilon_{r_l}>0$ design parameters as defined in (\ref{eq26}).

Finally we obtain
\begin{equation}
0\leq V_2(t)\leq \varepsilon/\lambda+\left[V_2(0)-\varepsilon/\lambda\right]e^{-\lambda t}. \label{eq28}
\end{equation}

\emph{Theorem 1}. Consider the trajectory tracking problem described by (\ref{eq8}) and (\ref{eq9}) under Assumption 1 and 2. If the control law is designed as in (\ref{eq25}), then the closed-loop tracking system is exponentially converge to a predefined compact set.

\emph{Remark 5}. Since $\epsilon_{u_l}$ and $\epsilon_{r_l}$ can be chosen arbitrarily small, the proposed tracking method guarantees exponential convergence of $p_e,~\psi_{le},~u_{le},~e_r$ to zero. In practice, however, excessively small values of $\epsilon_{u_l}$ and $\epsilon_{r_l}$ may result in high-gain control behavior, which can amplify noise and unmodeled dynamics. Therefore, these parameters should be chosen with appropriate caution during implementation.

\emph{Remark 6}. When $p_e=0$, we have $(u_{ld}^m,\psi_{ld}^m)=(u_{ld},\psi_{ld})$, and (\ref{eq25}) reduces to
\begin{align}
\begin{bmatrix}\tau_u\\ \tau_r\end{bmatrix}=&\bm{B}_l^{-1}\left\{\begin{bmatrix}\dot{u}_{ld}-f_{u_l}\\ \dot{\alpha}_{r_l}-f_{r_l}\end{bmatrix}+\bm{G}^{-1}\!\left(\bm{K}\bm{e}+\begin{bmatrix}0\\ \gamma_{\psi}\psi_{le}\end{bmatrix}\right)\right. \nonumber \\
&+\left.\begin{bmatrix}d_{u_lM}\varphi(\gamma_u u_{le}d_{u_lM})\\ d_{rM}\varphi(\gamma_r e_{r_l}d_{rM} )\end{bmatrix} \right\}, \label{eq29}
\end{align}
with $\alpha_{r_l}=\dot{\psi}_{ld}+\gamma_{\psi}^{-1}k_{\psi}\psi_{le}$. This shows that the singularity of $\psi_b$ at $p_e=0$ does not compromise the implementability of the proposed tracking controller $\bm{\tau}$ as in (\ref{eq25}).

\section{Surge Speed Positivity Enforcement}
In this section, the assumption $u(t)>0$ for all $t\geq0$ imposed in \cite{b6},\cite{b21}--\hspace{1sp}\cite{b24} will be relaxed in two different ways.

\subsection{Replacement by Relaxed Condition}
For most of practical marine vessels, limited propeller thrust and the constrained rudder size inherently restrict both the maximum forward speed and yaw rate. These performance limits can be determined through appropriate preliminary testing (e.t., VMT: vehicle maneuvering test) and regarded as part of the vessel's specifications \cite{b28}. Given these bounds on the forward speed and yaw rate, along with the vessel's passive-boundedness property \cite{b10}, it follows that the sway speed is also inherently bounded, and its maximum value can be determined in advance. Consequently, the following assumptin can be seen as highly reasonable in practical applications.

\emph{Assumption 3}. The vessel's sway velocity is bounded as $|v(t)|\leq v_M$, for all $t\geq0$ with $v_M>0$ a known constant.

\emph{Lemma 3:} If the design parameter $u_m$ in Assumption 2 is chosen as following
\begin{equation}
u_m>v_M+c_ua_p\sqrt{\varepsilon/c}+a_u\sqrt{\varepsilon/k_u}, \label{eq30}
\end{equation}
where $a_p,a_u>1$ are design parameters, then there exists a time $t_c$ such that $u(t)>0$, for all $t\geq t_c$.

\emph{Proof:} From (\ref{eq27}), it follows that there exists a time $t_c>0$ such that for all $t\geq t_c$, we have $p_e(t)\leq a_p\sqrt{\epsilon/k_p}$ and $|u_{le}|\leq a_u\sqrt{\epsilon/k_u}$. Moreover, since
\begin{align}
u_{ld}-u_{ld}^m&\leq c_up_e\leq c_ua_p\sqrt{\varepsilon/c}, \label{eq31}\\
u_{ld}^m-u_l&\leq a_u\sqrt{\varepsilon/k_u}, \label{eq32}
\end{align}
we have
\begin{equation}
u_{ld}-u_l\leq c_ua_p\sqrt{\varepsilon/c}+a_u\sqrt{\varepsilon/k_u}. \label{eq33}
\end{equation}

Consequently, we get
\begin{equation}
u_l\geq u_{ld}-c_ua_p\sqrt{\varepsilon/c}-a_u\sqrt{\varepsilon/k_u}. \label{eq34}
\end{equation}

Therefore, if the condition in (\ref{eq30}) holds, it follows from (\ref{eq34}) that: $\sqrt{u^2+v_M^2}\geq u_l>v_M$, which implies that $u(t)>0$, for all $t\geq t_c$. This completes the proof.
\qed

\emph{Remark 7}. It can be shown that, by selecting sufficiently large values of the design parameters $a_p$ and $a_u$, the time constant $t_c$ can be made arbitrarily small. Moreover, since the parameter $\varepsilon$ can be chosen arbitrarily small, at least theoretically, (\ref{eq30}) allows the constraint on the design of $u_{ld}(t)$ to be minimized.

\emph{Remark 8}. In implementation, for $t\in[0,t_c)$, if $u(t)=0$, it may be replaced by $u(t)=\delta_u$, where $\delta_u>0$ is a random value satisfying $\delta_u\leq \Delta_u$. Here, $\Delta_u>0$ denotes the upper bound on the measurement noise in the surge velocity, which can be determined based on the sensor specifications in practical applications. It is worth to mention that this replacement does not compromise the stability of the proposed closed-loop control system, since the effect of measurement noise has already been incorporated into the uncertainty term $d_{u_l}$ in (\ref{eq9}). On the other hand, when $u(t)>0$, the sideslip angle $\psi_a(t)$ is well defined in the domain $(-\pi/2,\pi/2)$. Consequently, both $b_{u_l}=\cos\psi_a b_u$ and $b_r$ are nonzero, ensuring that $\bm{B}_l$ in (\ref{eq21}) remains invertible regardless of whether $\epsilon_{ra}$ is zero or nonzero. i.e., whether the system is minimum-phase or not.

\subsection{Removal via a CBF-Based Technique}
Initially, CBFs were primarily studied in the context of system safety \cite{b25}--\hspace{1sp}\cite{b27}; more recently, they have been widely employed for singularity avoidance \cite{b32,b33}. With appropriately designed CBFs, various practical singularities can be effectively prevented. Nevertheless, as in controller design, uncertainty in the system dynamics complicates the construction of CBFs, which constitutes a common limitation of CBF-based approaches. In most cases, handling such uncertainty requires imposing additional constraints \cite{b34,b35}, and recent studies have further explored data-driven techniques for uncertainty estimation \cite{b36}. Since the primary contribution of this paper lies in the design of an EMO-based tracking controller, the problem is therefore addressed, for clarity and ease of discussion, under a generalized assumption -- namely, Assumption 1 -- in which the uncertainties are bounded by known constants.

The assumption $u(t)>0$ for all $t\geq0$ can be represented by the following CBF
\begin{equation}
h(\bm{\nu}_l)=u-\delta, \label{eq35}
\end{equation}
where $\delta>0$ indicates a constant design parameter specifying the safety margin from singularities.

For this barrier function in (\ref{eq35}) has relative degree 1, corresponding CBF is chosen as following \cite{b25}--\hspace{1sp}\cite{b27}
\begin{equation}
\dot{h}(\bm{\nu}_l,\dot{\bm{\nu}}_l)\geq-\alpha(h(\bm{\nu}_l)), \label{eq36}
\end{equation}
where $\alpha(\cdot)$ denotes any class-$\mathcal{K}$ function.

\emph{Proposition 2}. Barrier function condition for $u(t)>0$ for all $t\geq0$ can be presented as follows:
\begin{equation}
\tau_u\geq\left[d_{uM}-f_u-\alpha(u-\delta)\right]/b_u. \label{eq37}
\end{equation}

\emph{Proof}. For a given $\alpha(u-\delta)$, one can always design $\tau_u$ to satisfy (\ref{eq37}), which guarantees
\begin{align}
\dot{h}=&\dot{u}=f_u+b_u\tau_u+d_u\geq d_{uM}+d_u-\alpha(h(\bm{\nu}_l)) \nonumber \\
\geq& -\alpha(h(\bm{\nu}_l)). \label{eq38}
\end{align}
Satisfaction of (\ref{eq38}) further ensures that the barrier function $h(\bm{\nu}_l)$ defined in (\ref{eq35}) qualifies as a CBF. This completes the proof. \qed

Constraint (\ref{eq37}) can be expressed in the standard affine inequality form:
\begin{equation}
\bm{A}\cdot(\bm{\tau}-\bm{\tau}^*)\leq b, \label{eq39}
\end{equation}
where $\bm{A}=[-1~0]$, $\bm{\tau}^*$ denotes the reference control input, and $b=[f_u+\alpha(u-\delta)-d_{uM}]/b_u+\tau_u^*$.

As a result, the trajectory tracking problem considered in this paper can be formulated as the following convex quadratic program (QP):
\begin{align}
\underset{\dot{\bm{\nu}}_l,\bm{\tau}}{\text{min}}~~&||\bm{\tau}-\bm{\tau}^*||^2 \label{eq40} \\
\text{s.t.}~~&\text{Constraint (\ref{eq39})}, \nonumber
\end{align}
where $\dot{\bm{\nu}}_l=[u_l,r_l]^T$, and the reference control input $\bm{\tau}^*$ is computed according to (\ref{eq25}).

\emph{Assumption 4}. The control law $\bm{\tau}$ generated by QP (\ref{eq40}) is locally Lipschitz.

This standard assumption guarantees the existence and uniqueness of solutions to the closed-loop system, which is a fundamental prerequisite for establishing forward invariance of the safe set \cite{b26,b27}.

Consequently, by employing a CBF-based technique, the restrictive assumption $u(t)>0$, for all $t\geq0$ can be removed. Nonetheless, this approach remains subject to several limitations, including conservative behavior, feasibility issues, and high computational burden. In particular, when the system dynamics involve unknown uncertainties, the method lacks systematic design procedures.

\section{Numerical Studies}
This section demonstrates the proposed tracking scheme using MATLAB simulations, considering the dynamics of an underactuated marine vessel \cite{b14} together with (\ref{eq2}):
\begin{align}
f_u&=\left[m_{22}vr-\chi_{u1}u-\chi_{u2}|u|u-\chi_{u3}u^3\right]/m_{11}, \nonumber \\
f_v&=-\left[m_{11}ur+\chi_{v1}v-\chi_{v2}|v|v-\chi_{v3}v^3\right]/m_{22}, \nonumber \\
f_r&=\left[(m_{11}-m_{22})ur-\chi_{r1}r-\chi_{r2}|r|r-\chi_{r3}r^3\right]/m_{33}, \nonumber \\
b_u&=1/m_{11},~b_r=1/m_{33}, \label{eq41}
\end{align}
where $m_{11}=1.2e+5$, $m_{22}=1.779e+5$, $m_{33}=6.36e+7$, $\chi_{u1}=2.15e+4$, $\chi_{v1}=1.47e+5$, $\chi_{r1}=8.02e+6$, $\chi_{u2}=0.2\chi_{u1}$, $\chi_{u3}=0.1\chi_{u1}$, $\chi_{v2}=0.2\chi_{v1}$, $\chi_{v3}=0.1\chi_{v1}$, $\chi_{r2}=0.2\chi_{r1}$, $\chi_{r3}=0.1\chi_{r1}$. The uncertainty terms are assumed to be $d_u=22(1-2rand)/11$, $d_v=52(1-2rand)/17.79$, $d_r=190(1-2rand)/63.6$, where $rand\in[0,1]$ denotes uniformly distributed random noise. The lift effect is neglected by setting $\epsilon_r=0$. Accordingly, the uncertainty bounds are chosen as $d_{uM}=22/11$, $d_{vM}=52/17.79$, $d_{rM}=190/63.6$, and $d_{u_lM}=\sqrt{d_{uM}^2+d_{vM}^2}=3.542$.

The smooth reference trajectory is defined as follows: for $t\in[0,60s]$, $u_{ld}(t)=10m/s$ and $\psi_{ld}(t)=90deg$, with $\bm{\eta}_d(0)=[100m,30m,90deg]^T$; for $t\in(60s,75s]$, $u_{ld}(t)=10m/s$ and $\dot{\psi}_{ld}(t)=-0.05\exp[(t-75)/(t-60)]rad/s$; for $t>75s$, $u_{ld}(t)=10m/s$ and $\dot{\psi}_{ld}(t)=-0.05rad/s$. The vessel's initial conditions are set as $\bm{\eta}(0)=[50m,5m,30deg]^T$ and $\bm{\nu}(0)=[1m/s,0m/s,0rad/s]^T$, the controller design parameters are chosen as $k_{\psi}=40$, $k_u=800$, $k_r=100$, $\gamma_{\psi}=120$, $\gamma_u=60$, $\gamma_r=1$, $c_u=0.2$, $c_{\psi}=0.2$. The smooth function $\varphi(\cdot)$ in (\ref{eq26}) is selected as $\varphi(\zeta)=tanh(\sigma\zeta/\epsilon)$ with $\sigma=\epsilon=\epsilon_{u_l}=\epsilon_{r_l}=1$. Furthermore, the CBF related parameters are set to $\delta=0.6$, and the class-$\mathcal{K}$ function $\alpha(\cdot)$ is chosen as $\alpha(x)=x^2$.

As noted earlier, this paper proposes an EMO-based trajectory tracking scheme for underactuated marine vessels within a polar coordinate framework and investigates two relaxation methods for the restrictive assumption $u(t)>0$ commonly adopted in the literature. For convenience, the approach based on the relaxed condition introduced in Section IV is referred to as Method-1, whereas the CBF-based technique is denoted as Method-2.

\begin{figure}[!t]
\centerline{\includegraphics[width=\columnwidth]{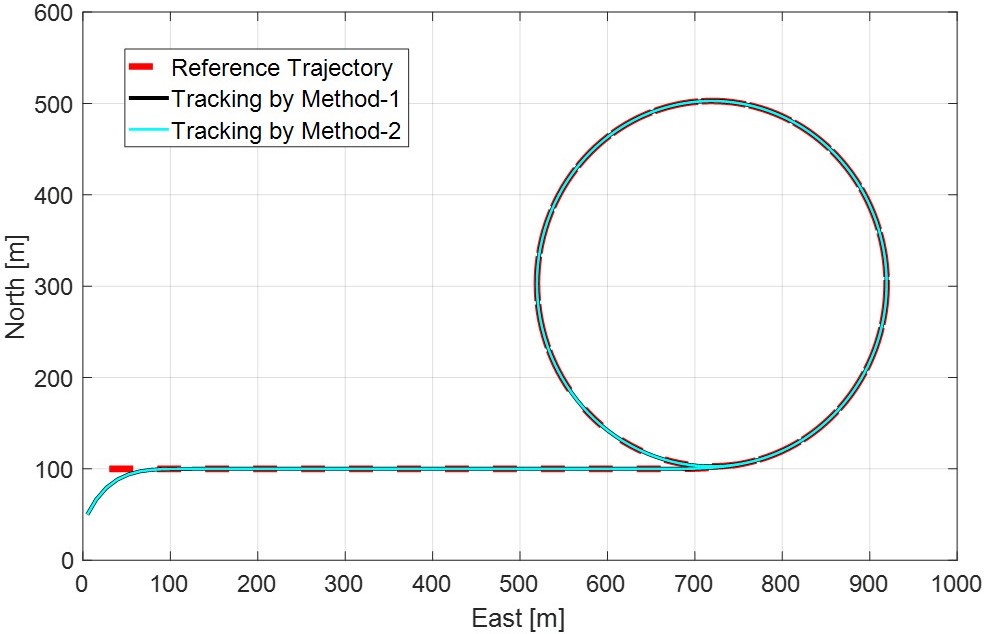}}
\caption{Reference trajectory and its tracking by two methods with $u_{ld}=10m/s$, $\forall t\geq0$.}
\label{fig2}
\end{figure}

\begin{figure}[!t]
\centerline{\includegraphics[width=\columnwidth]{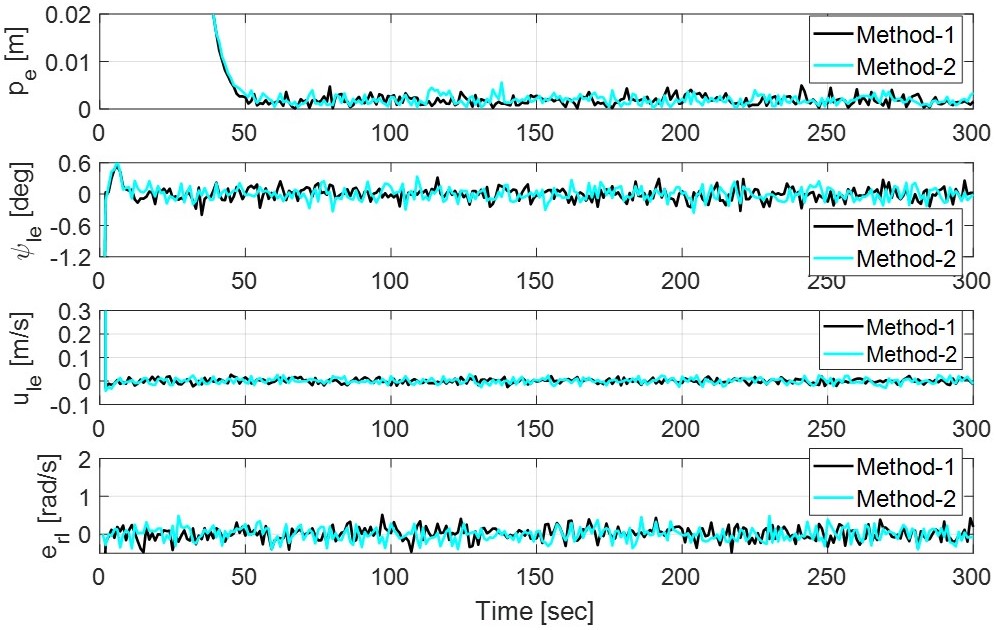}}
\caption{Tracking performance comparison with $u_{ld}=10m/s$, $\forall t\geq0$.}
\label{fig3}
\end{figure}

\begin{figure}[!t]
\centerline{\includegraphics[width=8.2cm]{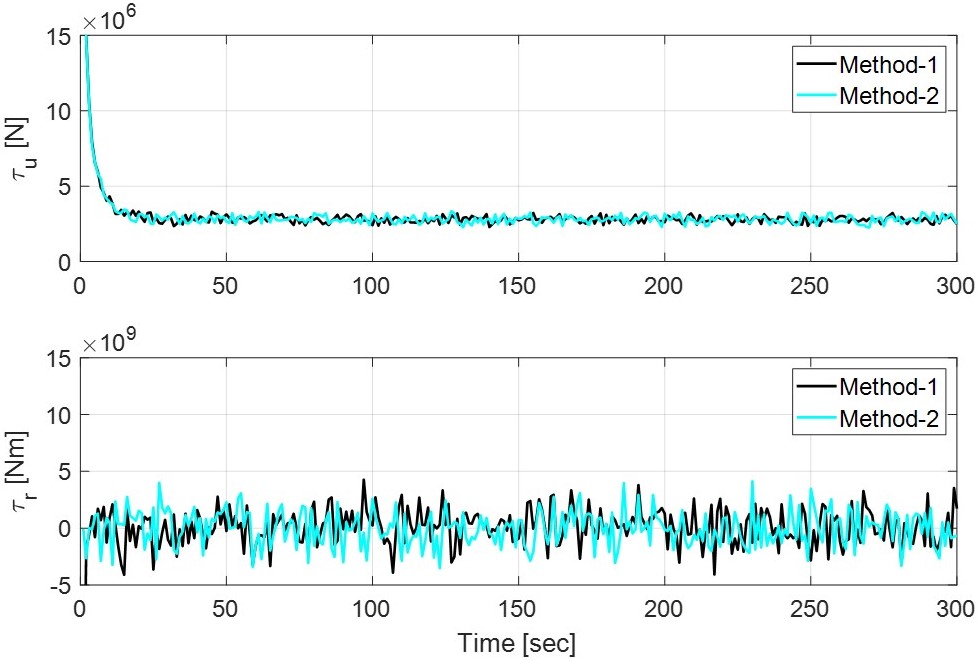}}
\caption{Corresponding control efforts comparison with $u_{ld}=10m/s$, $\forall t\geq0$.}
\label{fig4}
\end{figure}

Fig. \ref{fig2} illustrates the reference trajectory and the corresponding tracking performance of the two methods, which are identical in this case, indicating that the vessel motion induced by $\bm{\tau}^*$, computed according to (\ref{eq25}), satisfies constraint (\ref{eq38}). Fig. \ref{fig3} shows the tracking error convergence, and the corresponding control inputs are depicted in Fig. \ref{fig4}. The discrepancies observed in Fig. \ref{fig3} and \ref{fig4} arise from the use of different realizations of random noise.

\begin{figure}[!t]
\centerline{\includegraphics[width=\columnwidth]{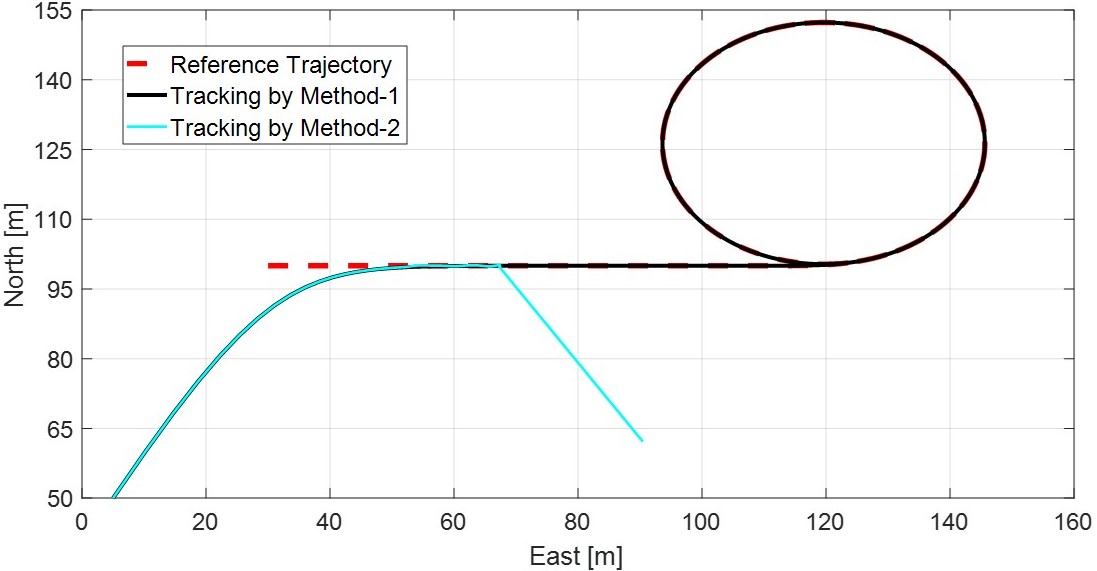}}
\caption{Trajectory tracking results in both methods with $u_{ld}=1.3m/s$, $\forall t\geq0$.}
\label{fig5}
\end{figure}

Compared with Method-2, Method-1 directly computes the control input as $\tau=\tau^*$, provided that constraint (\ref{eq30}) is satisfied. Assuming the maximum forward speed of the vessel to be $15m/s$ and the maximum yaw rate to be $0.05rad/s$ as specified above, it can be verified that the vessel’s sway velocity remains within $0.8m/s$. Given this value of $v_M$ and the associated parameters from the preceding settings, it isn straightforward to verify that the inequality (\ref{eq30}) holds for $u_m>1.4902m/s$. However, as illustrated in Fig. \ref{fig5}, Method-1 still achieves satisfactory tracking performance even when $u_m=1.3m/s$, which clearly violates constraint (\ref{eq30}). This observation indicates that condition (\ref{eq30}) is sufficient but not necessary.

\begin{figure}[!t]
\centerline{\includegraphics[width=\columnwidth]{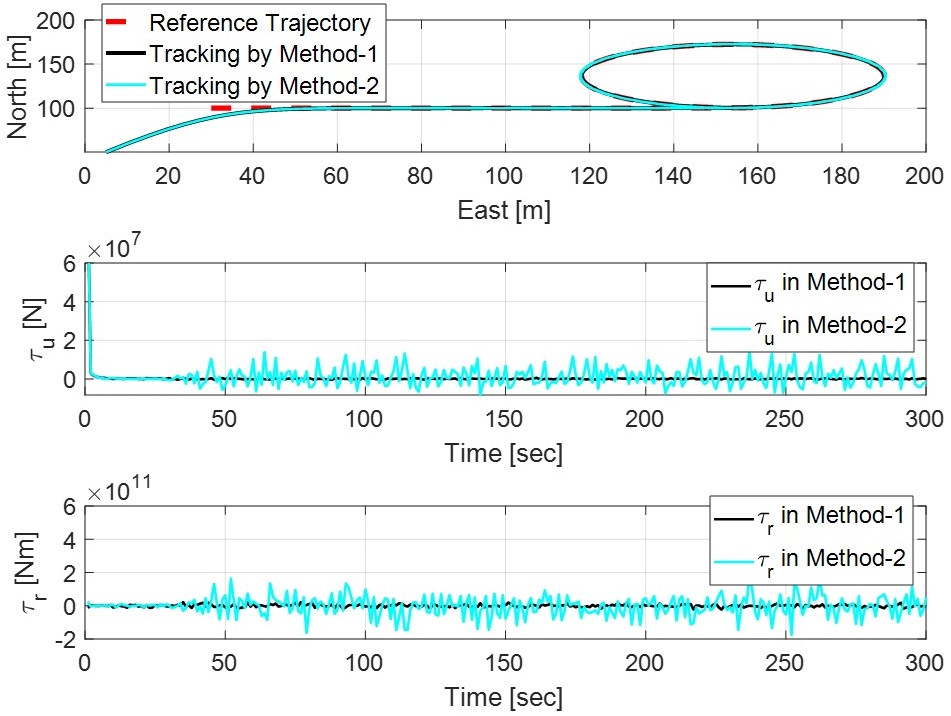}}
\caption{Trajectory tracking results and corresponding control efforts in both cases with $u_{ld}=1.8m/s$, $\forall t\geq0$.}
\label{fig6}
\end{figure}

An interesting observation is that, when $u_{ld}(t)$ is set to a small value, Method-1 is able to track the prescribed trajectory effectively, whereas Method-2 fails to do so, as illustrated in Fig. \ref{fig5}. This behavior persists until $u_m\geq 1.8m/s$. Fig. \ref{fig6} presents the tracking performance and corresponding control efforts of both methods for $u_{ld}(t)=1.8m/s$, from which it can be observed that Method-2 requires a higher control effort than Method-1.

In conclusion, the application of the proposed method indicates that the vessel’s reference trajectory is subject to certain constraints. Specifically, system stability is guaranteed only when the vessel’s reference forward speed exceeds a threshold value $u_m$, which is determined by the vessel specifications and operating conditions.

\section{Conclusion}
This paper presented a novel trajectory tracking scheme for underactuated marine vessels within a polar-coordinate framework. Two PCTs were employed to reformulate the original two-input–three-output tracking model into a two-input–two-output feedback form. To circumvent potential singularities arising in the recursive controller design, the EMO concept was incorporated. Furthermore, two alternative methods were proposed to relax the assumption of a strictly positive vessel surge speed, which had been originally imposed to address the singularity associated with the polar angle. The effectiveness of the proposed approach was demonstrated through numerical simulations.

It follows from the definition that the vessel’s azimuth angle is highly sensitive to noise when the position error is small. Accordingly, the design of an EMO scheme capable of attenuating this noise effect constitutes an important issue for future work.




\begin{thebibliography}{00}

\bibitem{b1} K. Y. Pettersen and H. Nijmeijer, ``Tracking control of an underactuated surface vessel,'' in \emph{Proc. of the 37th IEEE Conference on Decision \& Control}, Tampa, Florida USA, 1998, pp. 4561--4566.
\bibitem{b2} K. Y. Pettersen and H. Nijmeijer, ``Underactuated ship tracking control: theory and experiments,'' \emph{Int. J. of Control}, vol. 74, pp. 1435-1446, 2001.
\bibitem{b3} Z. P. Jiang, ``Global tracking control of underactuated ships by Lyapunov's direct method,'' \emph{Automatica}, vol. 38, pp. 301--309, 2002.
\bibitem{b4} K. D. Do, Z. P. Jiang, J. Pan, ``Underactuated ship global tracking under relaxed conditions,'' \emph{IEEE Transactions on Automatic Control}, vol. 47, pp. 1529--1536, 2002.
\bibitem{b5} T. C. Lee and Z. P. Jiang, ``New cascade approach for global $\kappa$-exponential tracking of underactuated ships,'' \emph{IEEE Transactions on Automatic Control}, vol. 49, pp. 2297--2303, 2004.
\bibitem{b6} J. H. Li, ``Path tracking of underactuated ships with general form of dynamics,'' \emph{Int. J. of Control}, vol. 89, pp. 506--517, 2016.
\bibitem{b7} H. Ashrafiuon, S. Nersesov, G. Clayton, ``Trajectory tracking control of planar underactuated vehicles,'' \emph{IEEE Transactions on Automatic Control}, vol. 62, pp. 1959--1965, 2017.
\bibitem{b8} J. M. Godhavn, ``Nonlinear tracking of underactuated surface vessels,'' in \emph{Proc. of the 35th IEEE Conference on Decision \& Control}, Kobe, Japan, 1996, pp. 975--980.
\bibitem{b9} E. Fredriksen and K. Y. Pettersen, ``Global $\kappa$-exponential way-point maneuvering of ships: Theory and experiments,'' \emph{Automatica}, vol. 42, pp. 677--687, 2006.
\bibitem{b10} J. H. Li, P. M. Lee, B. H. Jun, Y. K. Lim, ``Point-to-point navigation of underactuated ships,'' \emph{Automatica}, vol. 44, pp. 3201-3205, 2008.
\bibitem{b11} L. Consolini and M. Tosques, ``A minimum phase ouput in the exact tracking problem for the nonminimum phase underactuated surface ship,'' \emph{IEEE Transactions on Automatic Control}, vol. 57, pp. 3174--3180, 2012.
\bibitem{b12} A. P. Aguiar and J. P. Hespanha, ``Trajectory-tracking and path-following of underactuated autonomous vehicles with parametric modeling uncertainty,'' \emph{IEEE Transactions on Automatic Control}, vol. 52, pp. 1362-1379, 2007.
\bibitem{b13} C. Paliotta, E. Lefeber, K. Y. Pettersen, J. Pinto, M. Costa, J. Sousa, ``Trajectory tracking and path following for underactuated marine vehicles,'' \emph{IEEE Transactions on Control Systems Technology}, vol. 27, pp. 1423--1437, 2019.
\bibitem{b14} K. D. Do, Z. P. Jiang, J. Pan, ``Robust adaptive path following of underactuated ships,'' \emph{Automatica}, vol. 40, pp. 929--944, 2004.
\bibitem{b15} K. D. Do and J. Pan, ``Global robust adaptive path following of underactuated ships,'' \emph{Automatica}, vol. 42, pp. 1713--1722, 2006.
\bibitem{b16} T. I. Fossen and K. Y. Pettersen, ``On uniform semiglobal exponential stability (USGES) of proportional line-of-sight guidance laws,'' \emph{Automatica}, vol. 50, pp. 2912--2917, 2014.
\bibitem{b17} C. Samson, ``Path following and time-varying feedback stabilization of a wheeled mobile robot,'' in \emph{Proc. of ICARCV'92}, Singapore, 1992.
\bibitem{b18} N. Hung, F. Rego, J. Quintas, J. Cruz, M. Jacinto, D. Souto, A. Potes, L. Sebastiao, A. Pascoal, ``A review of path following control strategies for autonomous robotic vehicles: Theory, simulations, and experiments,'' \emph{Journal of Field Robotics}, vol. 40, pp. 747--779, 2023.
\bibitem{b19} C. Dong, B. Zheng, L. Chen, ``Trajectory tracking control for uncertain underactuated surface vessels with guaranteed prescribed performance under stochastic disturbances,'' \emph{Nonlinear Dynamics}, vol. 112, pp. 13215--13231, 2024.
\bibitem{b20} M. Krstic, I. Kanellakopoulos, P. Kokotovic. {\em Nonlinear and Adaptive Control Design}. John Wiley {\&} Sons, Inc., New York, 1995.
\bibitem{b21} J. H. Li, H. Kang, M. G. Kim, M. J. Lee, G. R. Cho, ``Asymptotic trajectory tracking of underactuated non-minimum phase marine vessels,'' \emph{IFAC PapersOnLine}, vol. 55, pp. 281--286, 2022.
\bibitem{b22} J. H. Li, H. Kang, M. G. Kim, H. S. Jin, M. J. Lee, G. R. Cho, ``Trajectory tracking performance transition analysis from ploar to Cartesian coordinates,'' \emph{IFAC PapersOnLine}, vol. 56, pp. 11615--11620, 2023.
\bibitem{b23} J. H. Li, ``3D trajectory tracking of underactuated non-minimum phase underwater vehicles,'' \emph{Automatica}, vol. 155, 111149, 2023.
\bibitem{b24} J. H. Li, H. Kang, M. G. Kim, M. J. Lee, H. S. Jin, G. R. Cho, ''Twoing type of 3D trajectory tracking for a class of underactuated autonomous underwater vehicles,'' in \emph{Proc. of American Control Conference}, Denvor, CO, pp. 3770--3775, 2025.
\bibitem{b25} M. Z. Romdlony and B. Jayawardhana, ``Uniting control Lyapunov and control barrier functions,'' in \emph{Proc. of 53rd IEEE CDC}, Los Angeles, CA, 2014, pp. 2293--2298.
\bibitem{b26} A. D. Ames, X. Xu, J. W. Grizzle, P. Tabuada, ``Control Barrier Function Based Quadratic Programs for Safety Critical Systems,'' \emph{IEEE Trans. on Automatic Control}, vol. 62, no. 8, pp. 3861--3876, 2017.
\bibitem{b27} A. D. Ames, S. Coogan, M. Egerstedt, G. Notomista, K. Sreenath, P. Tabuada, ``Control Barrier Functions: Theory and Applications,'' in \emph{Proc. of 18th European Control Conference (ECC)}, Napoli, Italy, 2019, pp. 3420--3431.
\bibitem{b28} T. I. Fossen. {\em Handbook of Marine Craft Hydrodynamics and Motion Control}. John Wiley {\&} Sons. Ltd, 2011.
\bibitem{b29} J-J. E. Slotine and W. Li. {\em Applied Nonlinear Control}. Prentice-Hall Inc., New Jersey, 1991.
\bibitem{b30} J. N. Newman. {\em Marine Hydrodynamics}. The MIT Press, Cambridge, MA, 1977.
\bibitem{b31} M. M. Polycarpou, Stable adaptive neural control scheme for nonlinear systems, {\em IEEE Transactions on Automatic Control}, vol. 41, pp. 447--451, 1996.
\bibitem{b32} V. Kurtz, P. M. Wensing, H. Lin, ``Control Barrier Functions for Singularity Avoidance in Passivity-Based Manipulator Control,'' in \emph{Proc. of 60th IEEE Conference on Decision and Control (CDC)}, Austin, Texas, 2021, pp. 6125--6130.
\bibitem{b33} M. Wu, A. Rupenyan, B. Corves, ``Singularity-Avoidance Control of Robotic Systems with Model Mismatch and Actuator Constraints,'' in \emph{Proc. of 23rd European Control Conference (ECC)}, Thessaloniki, Greece, 2025, pp. 2545--2550.
\bibitem{b34} P. Zhao, Y. Mao, C. Tao, N. Hovakimyan, X. Wang, ``Adaptive Robust Quadratic Programs using Control Lyapunov and Barrier Functions,'' in \emph{Proc. of 59th IEEE Conference on Decision and Control (CDC)}, Jeju Island, Republic of Korea, 2020, pp. 3353--3358.
\bibitem{b35} E. Das and J-W. Burdick, ``Robust Control Barrier Functions Using Uncertainty Estimation With Application to Mobile Robots,'' \emph{IEEE Transactions on Automatic Control}, vol. 70, no. 7, pp. 4766--4773, 2025.
\bibitem{b36} Y. Wang and X. Xu, ``Adaptive safety-critical control for a class of nonlinear systems with parametric uncertainties: A control barrier function approach,'' \emph{Systems\&Control Letters}, vol. 188, 105798, 2024.

\balance
\end{thebibliography}
\end{document}